# Exsolution of oxygen impurity from diamond lattice and formation of pressurized $CO_2$-I precipitates

Andrei A. Shiryaev [a,b,*], Yurii Chesnokov [c,d], Alexander L. Vasiliev [d,e], Thomas Hainschwang [f]

a) A.N. Frumkin Institute of Physical Chemistry and Electrochemistry, Russian Academy of Sciences, Leninsky pr. 31 korp. 4, 119071 Moscow, Russia

b) Institute of Ore Geology, Petrography, Mineralogy and Geochemistry (IGEM), Russian Academy of Sciences, Staromonetny per, 35, 119017 Moscow, Russia

c) National Research Center "Kurchatov Institute", Akademika Kurchatova pl., 1, 123182 Moscow, Russia

d) Shubnikov Institute of Crystallography of Federal Scientific Research Centre "Crystallography and Photonics" of Russian Academy of Sciences, Leninskii Prospekt 59, Moscow, 119333, Russian Federation

e) Moscow Institute of Physics and Technology, 9 Institutskiy Per., Dolgoprudny, Moscow Region, 141701, Russian Federation,

f) GGTL - GEMLAB Laboratory, Gnetsch, 42, LI – 9496 Balzers, Liechtenstein

*Corresponding author: shiryaev@phyche.ac.ru; a_shiryaev@mail.ru

**Abstract**

Diamond single crystals showing Infra-red features of pressurized $CO_2$-I phase were studied using Transmission Electron Microscopy (TEM) and tomography. Numerous O-containing precipitates with sizes up to 45 nm are observed. The absolute majority of these precipitates decorate dislocation loops or are located inside them; individual scattered precipitates are also present. Morphology of the precipitates varies from quasi-isometric octahedra to highly flattened elongated ones. Close association of the precipitates with the dislocation loops implies their formation by exsolution of oxygen impurity from diamond lattice; the size distribution of the precipitates suggests that the equilibrium state is not yet reached. Presumably, the morphology and precise chemical composition depend on P-T-t evolution of the diamond crystal and corresponding changes in oxygen supersaturation in the lattice. The $CO_2$- containing diamonds often contain micron-sized hexagonal lamellar inclusions. TEM investigation of a lamellae reveals that it consists of high quality graphite showing partial epitaxial relation with comprising diamond. The gap between the graphite and diamond may be enriched with oxygen impurity.

## 1. Introduction

Despite years of research, existence of oxygen as a lattice impurity in diamond remains debatable. Since early analyses using mass spectrometry [1] and nuclear probes [2,3], presence of oxygen was established in several diamond samples in concentrations up to 1000 at.ppm. However, since oxygen is an important constituent of most minerals and fluids, contribution of corresponding inclusions, possibly nanosized, is extremely difficult to rule out in these analytical methods. Computational studies [4-6] suggest that oxygen may form point defects in diamond lattice and spectroscopy provides some support for this hypothesis. Obviously oxygen ions may be driven into diamond lattice by ion implantation. Small amounts of luminescent [7], paramagnetic [8] and, possibly, electrically active [9] O-related defects were formed during post-irradiation annealing. For the purpose of the present work, it is more important to consider incorporation of O ions during diamond growth or during plasma treatment. Oxygen is believed to enter defects responsible for an optical absorption band with a peak at 480 nm [10], a peak at 1060 $cm^{-1}$ in infra-red (IR) [11,12], the 566 nm luminescence band [12] and several cathodoluminescence features [13,14]; paramagnetic defects are also ascribed to this impurity [5,15,16]. Despite all these studies, behavior of oxygen in diamond lattice is highly controversial and poorly understood.

Infra-red spectra of some natural diamond single crystals reveal absorption bands by $CO_2$ entities [17,18]. Significant variations in shape and position of corresponding spectral features both between samples and, sometimes, even within individual crystals were reported [10,18]. Recent detailed study revealed that the changes in the $CO_2$ spectral features across individual samples follow regular patterns [19]. It was suggested that the observations are explained by gradual changes in shape and/or chemical composition of nanosized $CO_2$-based inclusions, presumably, trapped during diamond growth. In the current study, we extend the work [19] by Transmission Electron Microscopy investigation of samples extracted from different parts of the

$CO_2$-containing diamond single crystals. The results of this work show that in the studied samples the $CO_2$-containing inclusions do not result from entrapment of a growth medium, instead, they were likely formed by decay of supersaturated solid solution of oxygen in diamond lattice. In addition, we report results of investigation of unusual type of graphitic inclusions encountered in these samples and rather common for $CO_2$-containing diamonds in general.

## 2. Materials and methods

Two natural diamond single crystals (FN7112 and FN7114) with strong $CO_2$-related bands in IR spectra were studied. The samples are described in detail in [19]. In short, they represent double-sided plates cut from gems and were treated at 6 GPa and 2100 °C for 10 minutes; the treatment did not influence the color and $CO_2$-related infra-red bands [10]. Despite being a single crystal, the sample FN7114 comprises two parts with markedly different coloration ("dark" and "transparent", see Fig. 1a); this zoning is also obvious in IR spectra. Using focused $Ga^+$ ion beam (FIB) thin foils were extracted from the "dark" and "transparent" parts of the specimen. Microstructural analyses of the FIB foils was performed using an OSIRIS TEM/STEM (Thermo Fisher Scientific, USA) equipped with a high angle annular (HAADF) electron detector (Fischione, USA) and Bruker energy-dispersive X-ray microanalysis system (Bruker, USA) at an accelerating voltage of 200 kV. Image processing was performed using a Digital Micrograph (Gatan, USA) and TIA (ThermoFisher Scientific, USA) software.

Both foils from the FN7114 specimen were also studied using TEM tomography to get information about 3D morphology of the precipitates. 10 nm gold nanoparticles (Colloidal Gold Labeled Protein A, UMC Utrecht, Netherlands) were used as fiducial markers for the alignment of tilt-series projection images. A 3 µL droplet of the suspension was applied on one side of the microscope grid, and then blotted to remove excess liquid and allowed to air dry. The tomographic study was carried out with a Titan Krios 60–300 TEM/STEM (FEI), equipped with TEM direct electron detector Falcon II (Thermo Fisher Scientific, USA) and Cs image corrector (CEOS, Heidelberg, Germany), at an accelerating voltage of 300 kV. Datasets were collected automatically with Tomography4 (Thermo Fisher Scientific, USA) software with 18000× magnification with the defocus value around – 0.5 µm by tilting the specimen within the −60° to 60° range with 1° step. Cross-correlation alignment and tomography restoration were performed using IMOD software [20] by weighted back-projection (WBP) method. Final visualization was made with UCSF Chimera [21] and IMOD. Note that tomographic reconstructions of the smallest precipitates is complicated by contribution of stress fields from dislocation loops and the selected threshold value is somewhat arbitrary.

An interesting feature of the studied diamond FN7112 is the presence of abundant small plate-like inclusions visible in optical microscope and special care was applied to extract an additional TEM foil containing such inclusion. Although the total number of the inclusions in the specimen is high, it is very difficult to find them in a thin (few microns) subsurface layer accessible for FIB sampling. The search for a suitable location was made using 3D optical microscope Keyence VHX-1000, which permits to make high quality images from a given depth in the sample. Scanning Electron microscopy showed that the inclusions exposed to the surface look like a thin crack. Upon examination of the polished diamond surface around the cracks, related bands with a negative surface relief may be traced, indicating that diamond was more efficiently removed by the polishing procedure immediately after the inclusion opening. The FIB foil was oriented perpendicularly to the plane of the platelet.

## 3. Results

### *3.1. Oxygen-containing precipitates*

Numerous nanosized inclusions were observed in the TEM foils. Their shapes and localization differ both between the samples and even in different parts of FN7114 diamond, where optically dark and bright parts are present (Fig. 1A).

In the optically dark part of the sample FN7114, as well as in the sample FN7112, numerous dislocation loops 50-100 nm in diameter are observed (Fig. 1). Many loops are decorated by precipitates (Fig. 1 and Supplementary Movies 3,4). The precipitates are also encountered inside the dislocation loops or may be present as scattered individual entities. In some cases a loop is not immediately obvious, but rather dense localization of the precipitates (e.g., in the upper right part of the Fig. 1C) likely indicate that the loop was initially present. Tomography rules out a hypothesis that the scattered "individual" inclusions are associated with loops oriented perpendicularly the observation plane in TEM images (Fig. 1H, Supplementary Fig. S1 and Movie SM1).

The tomographic slices shows rather wide scatter in shapes of the precipitates, which vary between almost perfect octahedra to elongated ones (Fig. 1 and Supplementary Movie 2),. The size distribution of the precipitates is shown in Fig. 1H and was calculated as (a+b)/2, where a, b – distances between the opposite (111) faces of a precipitate as observed in bright field TEM images (e.g., Fig. 1C). The error in the inferred size is mostly defined by uncertainty in determination of the precipitate edges in cases where strain-related contrast is sufficiently strong to blur the image. The overall uncertainty of the size is estimated as ~2 nm. Note that extraction of the size distribution from the tomographic images would gave much larger statistics, but the contrast issues definitely limits the precision, making the size determination more subjective. Most scattered "individual" precipitates are 7-10 nm in size; those associated with the dislocation loops are somewhat larger: the majority are between 7 and 15 nm. Diagonals of the largest precipitates reach 45 nm.

Appearance of the precipitates resemble N-containing voidites known in natural diamonds subjected to prolonged annealing. Voidites are often located within dislocation loops resulting from decomposition of platelet defects (e.g., [22,23]), or, in some diamonds, are unrelated to dislocations [24]. IR spectroscopy reveals low N content and absence of platelets in this part of the FN7114 specimen ([19], see also Figs. 1B and 3). Chemical mapping shows that the inclusions contain oxygen whereas nitrogen is not detected (Fig. 1M-O). In combination with infra-red spectroscopy data we consider the TEM observations as a strong support of a $CO_2$-based nature of the inclusions. In all cases the filling of the precipitates appears to be either amorphous or, possibly, poorly crystalline, since no hints for diffraction spots were found.

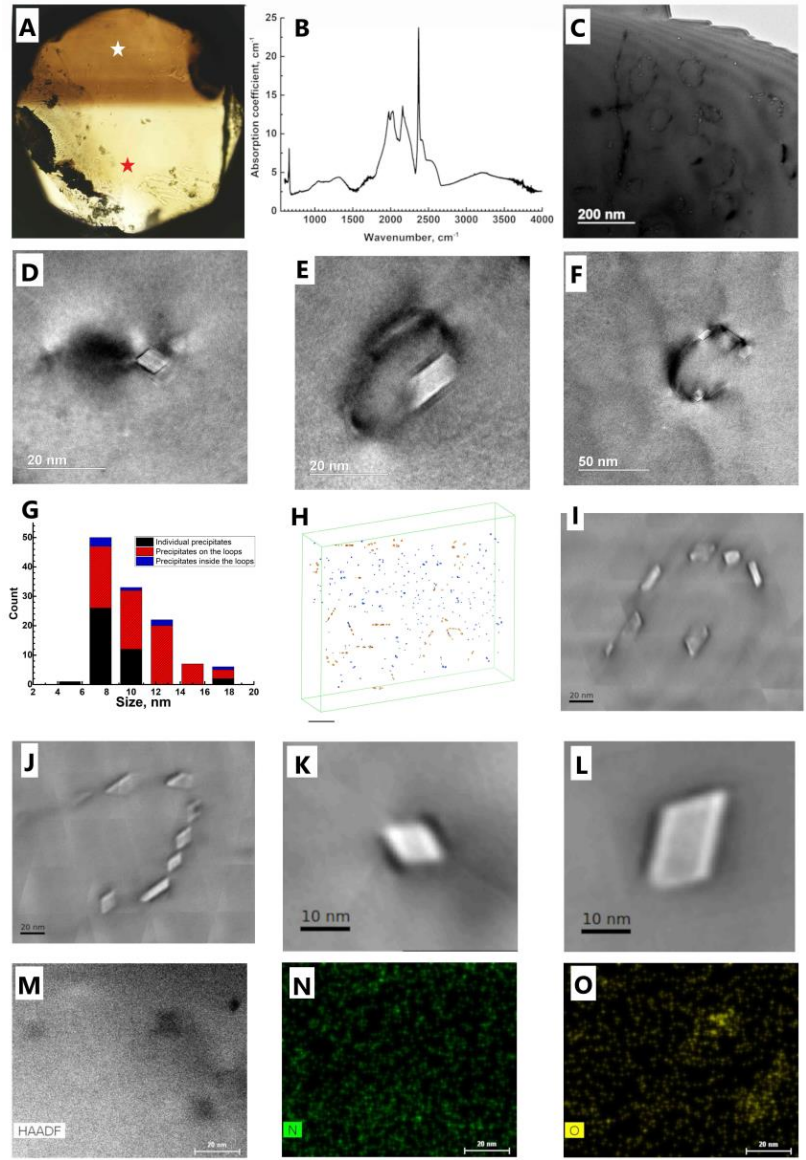

**Figure 1**. Electron microscopy and IR spectroscopy of the precipitates in dark part of the FN7114 (A-C, G-O) and in FN7112 (D-F) diamonds. A – optical photograph of the culet of the FN7114 sample; the field of view is 1.6 mm. White and red stars mark positions of FIB foils extraction in the dark and transparent parts, respectively. B - FTIR spectrum collected in the immediate vicinity of the TEM foil location. C – bright field TEM image showing dislocation loops with decorating precipitates. D - F – images of precipitates of quasi-isometric and elongated morphologies decorating dislocation loops. G – size distribution of the precipitates, see text for detail, H – electron tomography reconstruction of the FIB foil showing spatial distribution of the precipitates. Blue – individual precipitates; orange – those associated with dislocation loops. Scale bar – 200 nm. I-J - tomographic slices revealing different configurations of the precipitates on the loops. K, L - tomographic slices of individual precipitates with various elongation. M-O – HAADF image and maps of N and O distribution in a domain with the precipitates.

Another TEM foil was extracted from the optically bright transparent part of the diamond specimen (Fig. 2). Infra-red spectra of the corresponding location (Fig. 1D) indicate presence of N-containing platelet defects. Their peak is at 1365 cm$^{-1}$, corresponding to sizes ~40 nm [25]. Although complexity of one-phonon region of this crystal precludes detailed analysis, the presence of the platelets and of the B-defects suggest that the diamond is "regular" in Woods' classification [26] and N-related voidites are not expected to be present in substantial amounts. No platelets were observed in our TEM images, reflecting their low overall concentration. In this foil morphology of the $CO_2$-related inclusions markedly differ from those described above, see Fig. 2. The dislocation loops are present, but the main fraction of the decorating precipitates are not (quasi)isometric. Instead, most of them are represented by highly elongated and flattened octahedra with the long axis reaching 40-45 nm and with 4-8 nm smaller axes; size distribution of the precipitates is shown in Fig. 2G-I. The conclusions made from analysis of 2D images are fully supported by electron tomography (Fig. 2J,K, Supplementary Movies 5-7). Whereas chemical mapping in the foil from the optically dark part of the diamond (see above) showed clear correlation of the precipitates with oxygen impurity, in the "transparent" part of the specimen the concentration of this impurity is lower. Nevertheless, EDX line scans across the dislocation loops show presence of oxygen and absence of N (Fig. 2E,F).

Two foils were also extracted from the second diamond, FN7112. Whereas the principal attention was directed towards planar inclusions (see below), the O-related precipitates were also studied. Their morphology is generally close to isometric octahedra, although the elongated ones are also present (Fig. 1D-F).

The $CO_2$-related IR bands often show large variations in spectral envelope [10,18,19]. To explain these observations, two main hypotheses were proposed [19]: influence of a chemical impurity in the $CO_2$ ice and dependence of the envelope on morphology and internal structure (for example, core-shell) of the $CO_2$ ice nanoparticles. Comparison of IR spectra of regions analysed by TEM in both diamonds (Fig. 3) shows absence of obvious correlation between the $CO_2$-related IR absorption bands and morphology of the precipitates. Consequently, the scatter in IR manifestations of the $CO_2$ are more likely explained by the presence of a chemical impurity such as water, nitrogen or oxygen.

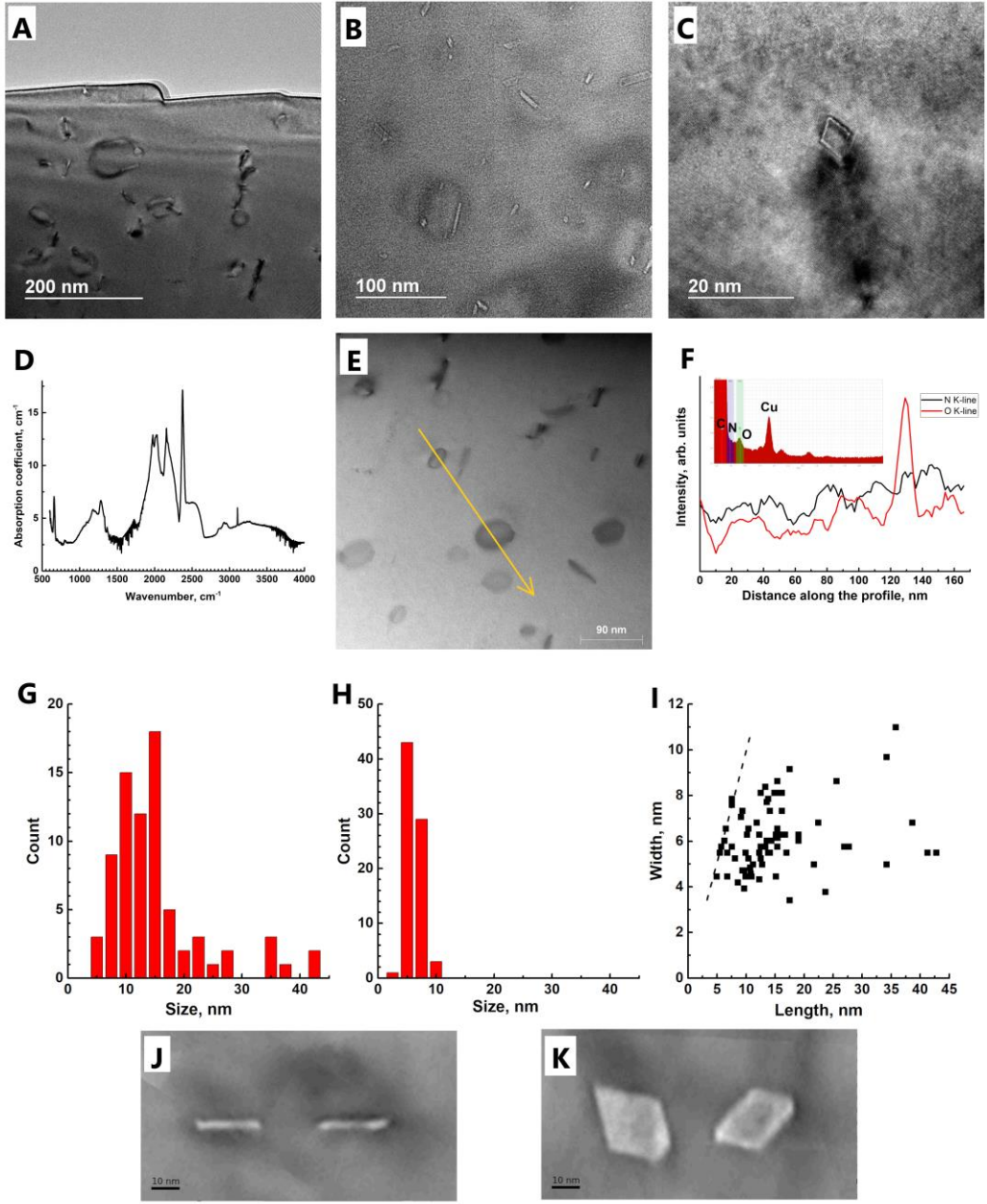

**Figure 2.** Electron microscopy and IR spectroscopy of the foil from the transparent part of the FN7114 diamond (see Fig. 1A for sample location). A-C – bright field TEM images showing dislocation loops with decorating precipitates and the isometric precipitate. D – IR spectrum collected in the immediate vicinity of the TEM foil location. E - low resolution image with indicated EDX profile (yellow line directed from left upper to lower right corner), F – intensity of K-lines of N and O along the profile; inset shows integral EDX spectrum. G, H – size distribution of precipitates' lengths and widths, respectively. I – length-width plot for the precipitates; dashed line show 1:1 ratio. J, K - tomographic slices showing two neighboring precipitates; the images are rotated relative to each other 90° around the horizontal axis to highlight flatness of the precipitates.

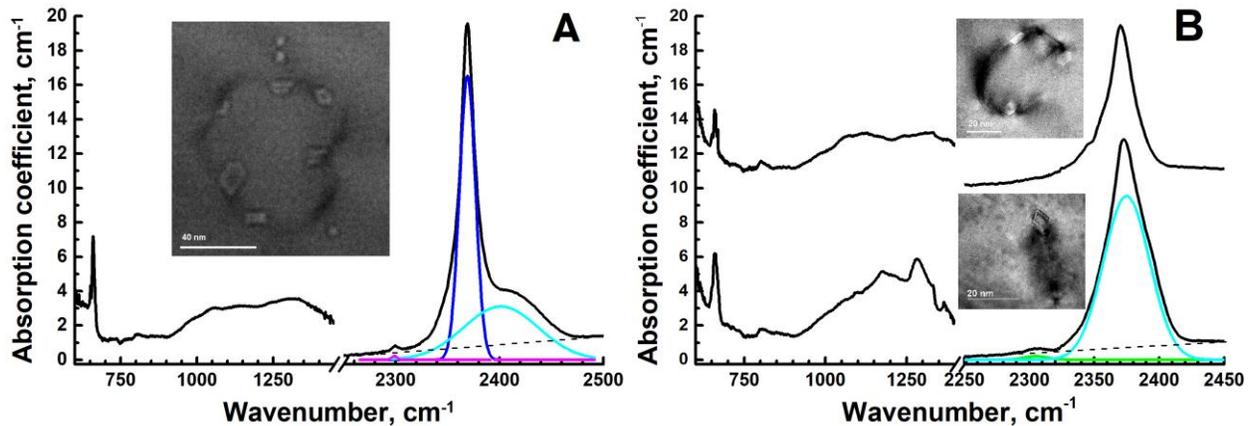

**Figure 3.** Infra-red spectra and TEM images of the samples FN7112 and FN7114. In IR spectra only regions of one-phonon absorption and of $CO_2$-related bands are shown. For the later, the contribution of diamond lattice absorption is subtracted to highlight the $CO_2$-bands; main components are shown (see also [19]). Dotted line is a linear background.

*3.2. Graphite platelets*

A remarkable feature of the studied samples and of $CO_2$-diamonds in general is the presence of small (5-10 µm) black (quasi)hexagonal inclusions with thickness below resolution of an optical microscope, i.e. less than ~0.5 μm [27,28]. Some of these inclusions apparently consist of several parallel hexagons (see Fig. 4A). Analysis of crystallographic relationships of the inclusions with the diamond matrix lead Lang and coauthors [27] to suggestion that the lamellas are graphitic. In the same time, our repeated attempts to detect Raman spectrum from the lamellae using various excitation wavelengths in confocal mode or to observe their X-ray diffraction pattern failed to identify the phase(s) present in the inclusions.

The TEM image of the FIB foil containing the lamellae in shown in Fig. 4. Electron diffraction and direct imaging of interlayer spacing indicate that the lamellae is indeed pure graphite (Fig. 4 G,H), comprising at least two parallel subdomains. Mottled contrast in some regions and streaks between the diffraction spots indicate disorder. The crystallographic orientation of the graphite-diamond is close to G[001]-D[111]. The contact between the graphitic sheet and the diamond matrix differs between the "upper" and "bottom" parts of the lamellae as shown in Fig. 4C-F. In the bottom part, the contact was likely (quasi)epitaxial. As discussed in detail in [29] the misfit in the interface G(001)/D(111) is only ~2% at normal conditions. The mismatch strain caused slight undulation of the graphite manifested as regular detachments between the two phases. In the upper part, the graphite lamellae is separated from the diamond matrix by a ~10 nm gap. Chemical mapping reveals presence of oxygen in the gap (Fig. 4I-L). It is unlikely to be an atmospheric contamination, since if this would be the case, oxygen would be present in many other spots, for example, at the contact of the graphitic lamellae with the diamond surface exposed by mechanical polishing (upper right corner of the image). Moreover, such a contamination should contain nitrogen, which is not observed. In our view, the oxygen precipitation in the gap may be one of the reasons for partial local graphitization of diamond or its consequence, see Discussion.

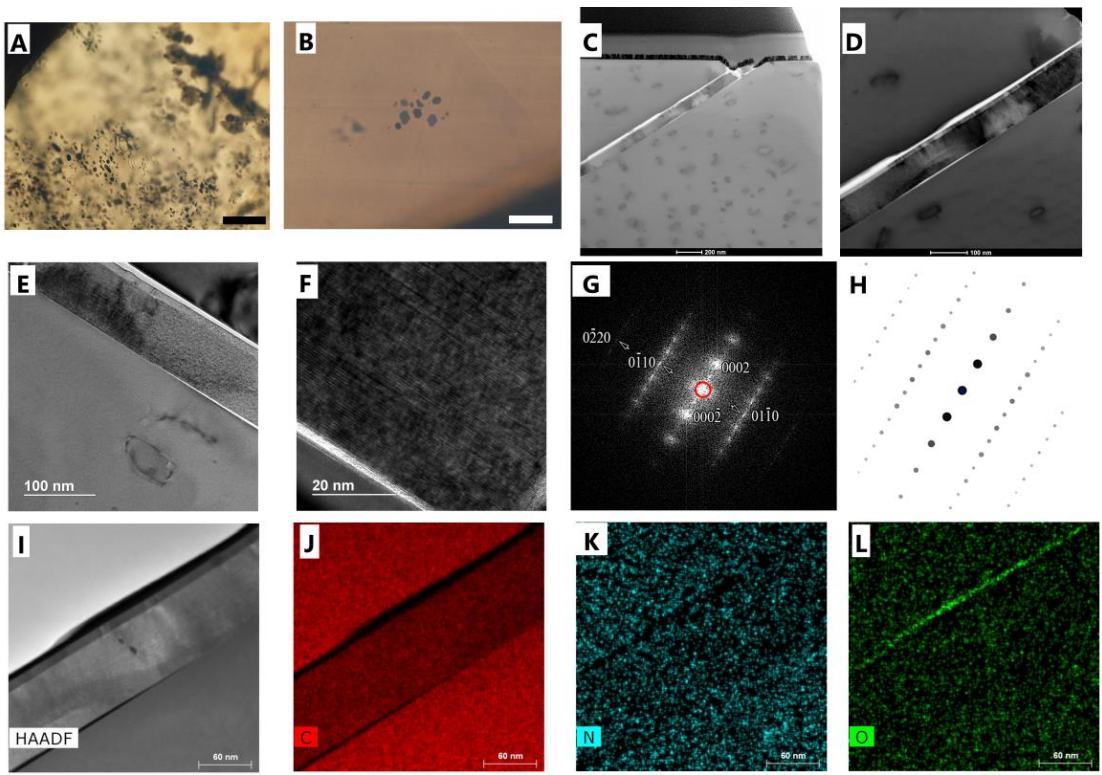

**Figure 4.** Graphitic platelets. The FN7112 diamond with abundant inclusions. A – Optical microphotograph of FN7112 showing numerous dark inclusions (dark feature in the right upper corner is a crack); field of view is 500 μm. Slightly below the image center stepped inclusion opened during polishing reveals that some of them likely comprised several parallel individs. B – a small cluster of graphitic platelets in the dark part of FN 7114; the scale bar is 50 μm. C-F – Bright field TEM images of the foil comprising the platelet-like inclusion cut normal to its plane. Numerous dislocation loops decorated with $CO_2$-precipitates are present. Note gaps between the graphitic platelet and diamond matrix. Dark layer on top in C – remnants of protective Pt coating used during foil preparation. E, F – Zoomed view of the graphitic platelet. The platelet consists of two parallel units. G, H – Fourier transform of F showing that the platelet is represented by disordered 2H graphite; H – theoretical pattern from the 2H graphite [30]. I-L – HAADF image and EDX maps of C, N, O, respectively.

## 4. Discussion.

### *4.1. Formation of the oxygen-containing precipitates*

Dislocation loops are not very common in as-grown diamond. Degradation of the N-containing platelets occurring due to insufficiently understood reasons is, perhaps, a most common process producing the loops in natural diamond. The former platelets are often marked by dislocation loops, frequently surroundung abundant nanosized precipitates of relatively low electronic density, called voidites. The voidites contain N- or N-H species, presumably inherited from the parent platelets; exact composition and degree of crystallinity appears to be sample-dependent. Sometimes voidites are strongly elongated [22], similar to precipitates in the transparent part of FN7114 (Fig. 2). As discussed above, the precipitates observed by us are generally larger than the voidites, contain oxygen instead of nitrogen and, in most cases, are directly attached to a dislocation loop rather than being dispersed inside. The precipitates are also unrelated to the N-containing platelets.

Inclusions of trapped $CO_2$ with sizes reaching several microns in size are known in natural [17, 18, 31-33] diamonds as well as in some synthetic single crystals [12]. There is little doubt that such inclusions were trapped during diamond growth or fill healed cracks. We suggest that the precipitates observed by us were formed by fundamentally different mechanism: exsolution of oxygen impurity present in diamond lattice. The formation of nanosized impurity precipitates decorating dislocation loops is a common phenomenon in supersaturated solid solutions (e.g., in heavily doped semiconductors) and in irradiated materials. Upon cooling, concentration of intrinsic and/or extrinsic defects may exceed respective solubility limits and thus precipitation of impurities and/or vacancies occurs. In these instances, formation of dislocation loops is a result of agglomeration of excess vacancies or interstitials. The loops can also be generated by a precipitate itself to decrease mechanical stresses, but in this situation, undecorated nested loops are commonly formed.

In our view, the only feasible explanation of all observations of the O-rich precipitates decorating dislocation loops is the decay of supersaturated solid solution of oxygen in diamond lattice. This hypothesis explicitly implies presence of an oxygen impurity in as-grown diamond. In silicon – a material with diamond lattice – the morphology of the $SiO_x$ precipitates depends both on exact precipitate composition and on supersaturation of oxygen in silicon lattice. Based on experimental observations and conception of equilibrium growth morphology, it was suggested that the $SiO_x$ precipitates form platelets at low temperatures and high supersaturations, and octahedra at high T's and low supersaturations [34]. Exact chemical composition, degree of crystallinity and internal structure of the precipitates also depend on temperature and on properties (impurities, type of conductivity, etc.) of the silicon sample. Similar trends could be expected for oxygen in diamond. Thermal history of the diamond sample FN7114 obtained from IR-active defects indicates that the dark part of the sample grew at lower temperatures, than the bright one. This scenario fits well the transition from the (roughly) octahedral $CO_2$ precipitates to platelets-like: the latter ones form in the cooler environments.

According to theory of precipitate formation in solid solutions [35], the asymptotic (reduced) size distribution of the precipitates does not depend on initial state of the system and a limiting value of a reduced size should be reached. Although it is possible that we underestimate population of the smallest precipitates, the overall shape of the size distribution (Figs. 1G and 2G) function clearly shows a pronounced tail towards larger precipitates. Assuming applicability of the model from [35], we conclude that the process of oxygen precipitation from diamond lattice has not yet reached a steady state. The presence of rare large precipitates may reflect interaction between them, which is especially plausible in view of high density of the precipitates decorating the loops.

Although total concentration of oxygen and nitrogen in diamond may be correlated [36], spectroscopically-active nitrogen is always anti-correlated with the IR-observable $CO_2$ [10,19]. Since the amount of the nitrogen and hydrogen impurities and their speciation change across the diamond specimen FN7114, it is plausible that exact chemical composition of the precipitates may vary both in different parts of heterogeneous diamonds (e.g., in dark and transparent parts of FN7114) and between individual samples. This hypothesis is supported by comparison of the TEM and IR data, see Fig. 3.

The formation of the $CO_2$-based precipitates by the exsolution of oxygen from diamond lattice has important implications for reconstruction of P-T history of a diamond specimen from pressure-induced shifts of IR or Raman bands. As discussed in [19], the observed evolution of the inferred residual pressure across diamond samples clearly indicate that the shift of spectral lines is not always confined to the pressure effects; morphology and exact chemical composition of the precipitates may also contribute with different signs. The current work adds another important parameter: if the precipitates were formed by the exsolution of impurity atoms, their composition, structure and morphology and, consequently, apparent shift of the spectral lines, depend on solubility and speciation of the relevant impurity in the crystalline lattice. These factors, in turn, depend on large number of poorly constrained parameters related to diamond growth medium composition and P-T-$fO_2$ conditions. As an extreme example, one may recall high residual pressures observed in crystalline precipitates of noble gases produced in ion-implanted materials, which were always kept at ambient pressure (e.g., [37]). Similar to the formation of $CO_2$-precipitates studies by us, voidites in diamond are produced by exsolution of excess N from the diamond lattice. Consequently, the shifts of $N_2$ [38] and $CO_2$-related IR/Raman bands as well observation of compression of the lattice fringes [39] in these precipitates are not necessarily applicable to depth of formation of comprising diamonds.

The $CO_2$-diamonds possess number of absorption bands in one-phonon region of IR spectra [10], Fig. 3. Whereas some of them are associated with silicates and carbonates (see maps in [19]), the bands peaked at ~980-990, 1050-1060, 1120-1140, ~1240, ~1300 $cm^{-1}$ remain unassigned. In all spectra of these diamonds, a sharp peak at diamond Raman frequency of 1332 $cm^{-1}$ is present. Its importance stems from the fact that local perturbation of diamond symmetry is a prerequisite of its appearance in the IR spectrum [40]. Whereas the assignment of these bands to a particular defect(s) is impossible at present, we tentatively suggest that the reported absorptions might be related to oxygen-related defect(s) in the diamond lattice.

### *4.2. Graphite platelets*

A remarkable feature of substantial number of $CO_2$-containing diamonds is the presence of tiny graphite lamellae in the crystal bulk. The concentration of these defects varies widely: whereas in the FN7112 they are filling the main fraction of the sample volume, in FN7114 they form very small scattered clusters (Fig. 4A,B). Based on a comprehensive review by Khokhraykov and Nechaev [41], we classify the inclusions as syngenetic. Although both studied crystals were HPHT annealed in the diamond stability field, no changes in visual appearance and/or spectroscopic features were observed. Moreover, the graphitic lamellae may be abundant even in untreated natural diamonds [27,28]. Consequently, their formation suggests growth of such crystals close to the diamond-graphite equilibrium line. As suggested in [42], heating of a mineral inclusion in diamond may lead not only to local cracking of the diamond due to differences in thermal expansion coefficients, but also may release some oxygen. This gas may, in turn, promote graphitization of the crack surfaces. In the case of the $CO_2$ diamonds, mineral inclusions are few. High magnification optical microscopy shows absence of a correlation between spatial distribution of the graphitic lamellae and the inclusions. In the studied sample FN7112, the gap between the graphite lamellae and the diamond matrix shows enrichment in oxygen. In close vicinity to the lamellae abundant O-containing nanosized precipitates are present. This coexistence suggests that formation of the $sp^2$-C phase and the precipitates were independent processes, but the gap may have trapped some of the diffusing oxygen atoms.

## 5. Conclusions

Transmission electron microscopy of diamond singe crystals manifesting strong $CO_2$-related absorption in IR spectra reveals presence of numerous O-rich nanosized precipitates; often associated with dislocation loops. Joint consideration of the current TEM results and detailed IR study [19] strongly suggest a direct link between the $CO_2$-related IR features and the precipitates. It is highly likely that both the O- and N-containing nanoprecipitates, the later known as voidites, were formed by a common mechanism: exsolution of oxygen (nitrogen) impurity from the diamond lattice. Variations of morphology of the precipitates – plate-like or octahedral - is likely governed by temperature of their formation. This model explicitly implies that in some diamonds oxygen impurity enters lattice sites. However, their spectral identification, if any, remains elusive.

The driving force for nucleation and subsequent growth of the precipitates is determined by difference between the impurity incorporation rate during crystal growth and its equilibrium solubility at given thermodynamic conditions. Both of these parameters depend on large number of parameters. Consequently, in the case of the $CO_2$-precipitates as well as in case of N-related voidites, the shift of spectral IR/Raman lines cannot be used for reconstruction of pressure of the diamond formation.

Detailed investigation of dark hexagonal plates common for $CO_2$-diamonds unambiguously proves their assignment to very thin graphite lamellae, most likely, of syngenetic origin. The gap between the graphite and enclosing diamond may contain oxygen impurity. However, at present it is unclear, whether this correlation is of genetic significance or is a coincidence.

**Acknowledgements.** We thank Drs. V. Artemov for preparation of FIB foils and V. Yapaskurt for assistance with SEM.

**References.**